\documentclass[useAMS,usenatbib]{mn2e}
\usepackage{graphicx}

\def\aV{\mbox{$\rm A_V$}}

\def\mMJ{\mbox{$(m-M)_J$}}
\def\mMo{\mbox{$(m-M)_O$}}
\def\ebv{\mbox{$E(B-V)$}}

\def\ejh{\mbox{$E(J-H)$}}

\def\rc{\mbox{$R_{\rm c}$}}

\def\rl{\mbox{$R_{\rm RDP}$}}

\def\ms{\mbox{$M_\odot$}}
\def\ds{\mbox{$d_\odot$}}

\def\dSC{\mbox{$\Delta R_{\rm SC}$}}

\def\jj{\mbox{$J$}}
\def\hh{\mbox{$H$}}
\def\ks{\mbox{$K_s$}}

\title[The embedded clusters related to the H\,II region NGC\,2174]
{Uniform detection of the pre-main sequence population in the 5 embedded 
clusters related to the H\,II region NGC\,2174 (Sh2-252)}

\author[C. Bonatto and E. Bica]{C. Bonatto$^1$ and E. Bica$^1$\\
$^1$ Departamento de Astronomia, Universidade Federal do Rio Grande 
do Sul, Av. Bento Gon\c{c}alves 9500\\
Porto Alegre 91501-970, RS, Brazil}

\begin{document}

\pagerange{\pageref{firstpage}--\pageref{lastpage}}

\maketitle

\label{firstpage}

\begin{abstract}
{\bf We investigate 5 embedded clusters (ECs) and the extended stellar group itself of the 
prominent H\,II region NGC\,2174 (Sh2-252), which presents scarce and heterogeneous information, 
coming from the optical and infrared. Considering the discrepant values of distance and age, 
the clusters and the H\,II region appear to be physically unrelated. The analysis is based 
on field-star decontaminated 2MASS photometry, which allows sampling the pre-main sequence 
(PMS). We find that Sh2-252A, C, E, NGC\,2175s, and Teu\,136 are small ECs (radius within 
$1.0 - 2.3$\,pc) characterised by a similar age ($\sim5$\,Myr), reddening ($\aV\sim1$), 
distance from the Sun ($\ds\sim1.4$\,kpc), and low mass ($60-200\,\ms$). This age is 
consistent with the H\,II region, the presence of O and B stars still in the MS, and 
the dominance ($\ga95\%$ in number) of PMS stars in colour-magnitude diagrams (CMDs). 
NGC\,2175 is not a star cluster, but an extended stellar group that encompasses 
the ECs Sh2-252\,A and C. It contains $\sim36\%$ of the member stars (essentially PMS) 
in the area, with the remaining belonging to the 2 ECs. CMDs of the overall star-forming 
region and the ECs provide $\ds=1.4\pm0.4$\,kpc for the NGC\,2174 complex, consistent with 
the value estimated for the physically-related association Gem\,OB1. Our uniform approach 
shows that NGC\,2174 and its related ECs (except, perhaps, for Teu\,136) are part of a single 
star-forming complex. CMD similarities among the ECs and the overall region suggest a coeval 
(to within $\pm5$\,Myr) star-forming event extending for several Myr. At least 4 ECs 
originated in the event, together with the off-cluster star formation that probably gave 
rise to the scattered stars of NGC\,2175. }
\end{abstract}


\section{Introduction}
\label{Intro}

{\bf A lingering question related to star formation is whether stars in associations and 
young stellar groups originate in clusters that dissolve shortly (e.g. \citealt{vdB92} 
and references therein), or are directly formed throughout the parent molecular cloud. 
At the bottom of this issue lies the scenario in which star formation is scale-free and
hierarchical, with high velocity turbulent gas forming large-scale structures and small 
clumps being formed by low-velocity compression (\citealt{ElmeCon}). In this context, 
young stellar groupings would be hierarchically clustered, with the great star complexes
at the largest scales and the OB associations and subgroups, clusters and cluster sub-clumps 
at the smallest (e.g. \citealt{Efremov95}). 


Hierarchical patterns in extended structures have been detected in the Magellanic Clouds (MCs) 
and other nearby galaxies (e.g. \citealt{2PCF}, and references therein). At the few parsec 
scale, Hubble Space Telescope (HST) and VLT-ISAAC photometry of pre-main sequence (PMS) stars 
in the SMC star-forming region NGC\,346/N\,66 suggest hierarchical star formation, probably
originated from more than one event (\citealt{Hennekemper08}; \citealt{Gouliermis08};
\citealt{Sabbi07}). These works show that part of the PMS stars are found in sub-clusters 
(some located in the central region of the association and others at the border), with the
remaining stars scattered around the association, a scenario directly related to the single 
or sequential star formation issue.


Recently, our group studied the stellar content of the Sh2-132 H\,II region, a star-forming 
complex hosting at least 4 embedded clusters (ECs) and presenting evidence of triggered star 
formation and hierarchical structuring (\citealt{Sh2-132}). However, given its rather large 
distance ($\ds\sim3.5$\,kpc) and the 2MASS\footnote{The Two Micron All Sky Survey, All Sky 
data release (\citealt{2mass06}).} photometric depth, we could only sample a fraction of the 
PMS population. Thus, the present paper focuses on the nearby H\,II region NGC\,2174 (Sh2-252)
and related ECs. This complex is located in the 
Orion constellation and appears to be physically related to the Gem\,OB1 association
($\ds=1.5-1.9$\,kpc, \citealt{Dunham2010}), together with several other Sharpless H\,II regions 
and the supernova remnant IC\,443 (see \citealt{Dunham2010} for a discussion of the objects 
in the area). Star cluster studies in this region are rather scarce and not uniform, presenting 
very different values of distance and age, to the point that no physical relation among the 
clusters and the complex appears to exist (Sect.~\ref{RecAdd}). Thus, our main goal is to 
investigate the relation of the complex and star clusters. To do so, we must go deep into the 
PMS, and work both with MS and PMS isochrones to characterise such clusters in detail, and 
derive accurate CMD fundamental and structural parameters. }

\begin{figure}
\resizebox{\hsize}{!}{\includegraphics{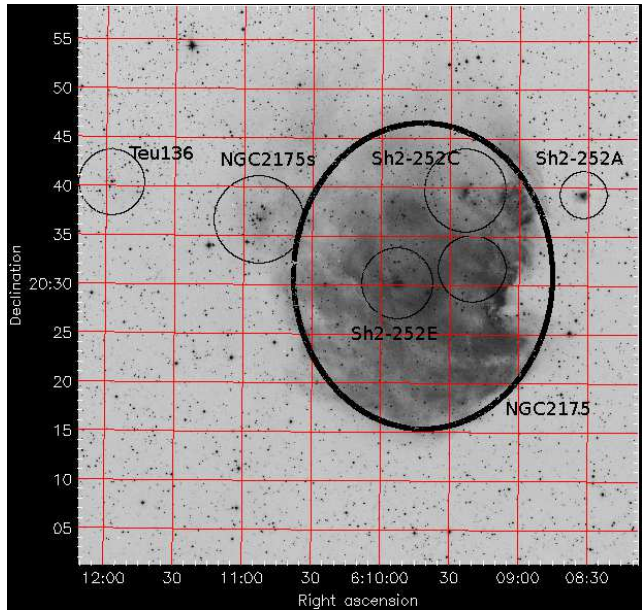}}
\caption[]{$45\arcmin\times45\arcmin$ DSS-I B image of the H\,II region and related
clusters. The thick circle marks off the area encompassed by the stellar group NGC\,2175, 
according to the literature. The small unlabelled circle indicates an inner region of 
NGC\,2175 that is free from contamination by stars of Sh2-252C and Sh2-252E. Gas emission, 
dust reflection and/or absorption are present in the field in varying proportions. North 
to the top and East to the left.}
\label{fig1}
\end{figure}

\begin{figure}
\begin{minipage}[b]{0.50\linewidth}
\includegraphics[width=\textwidth]{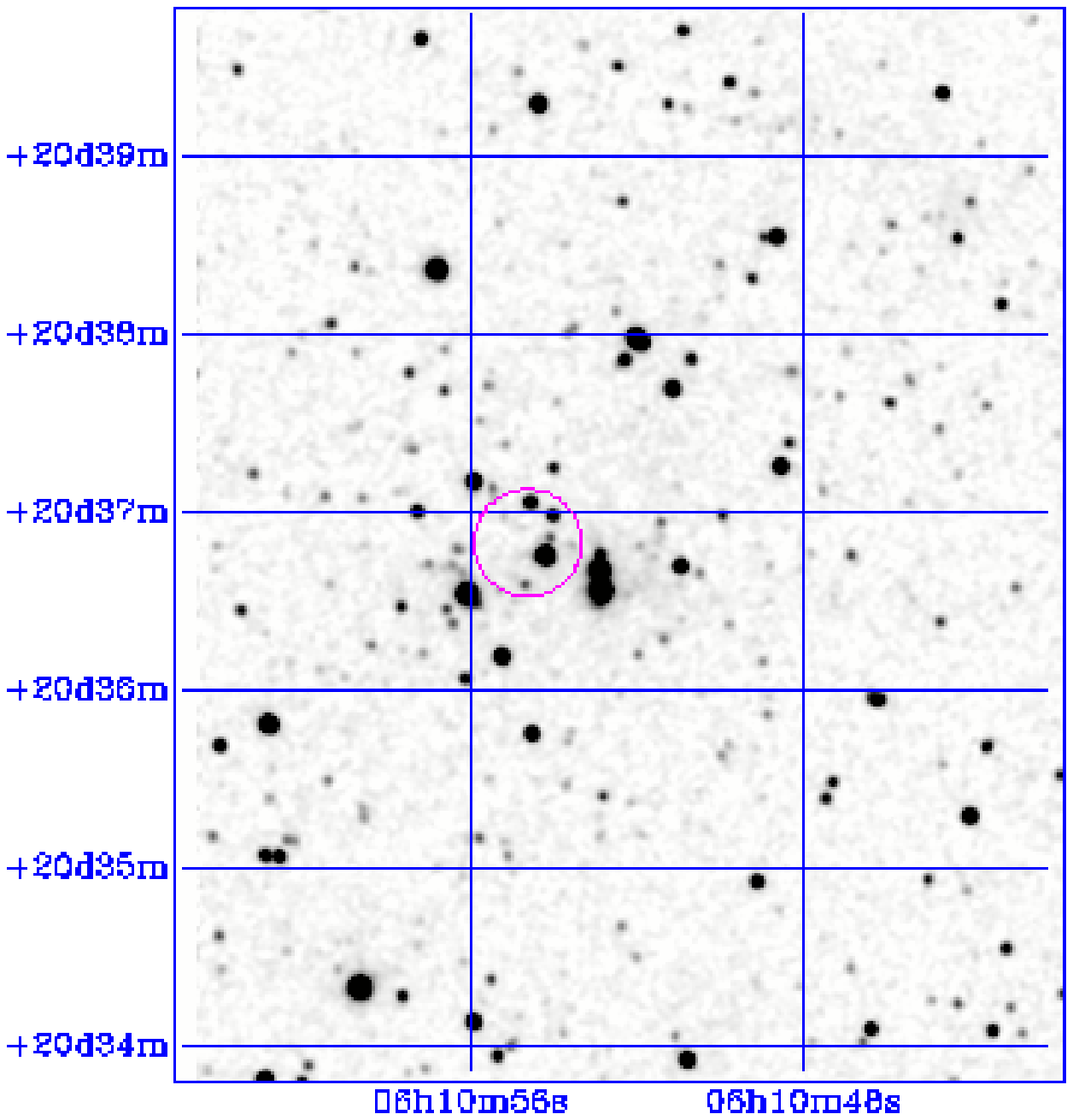}
\end{minipage}\hfill
\begin{minipage}[b]{0.50\linewidth}
\includegraphics[width=\textwidth]{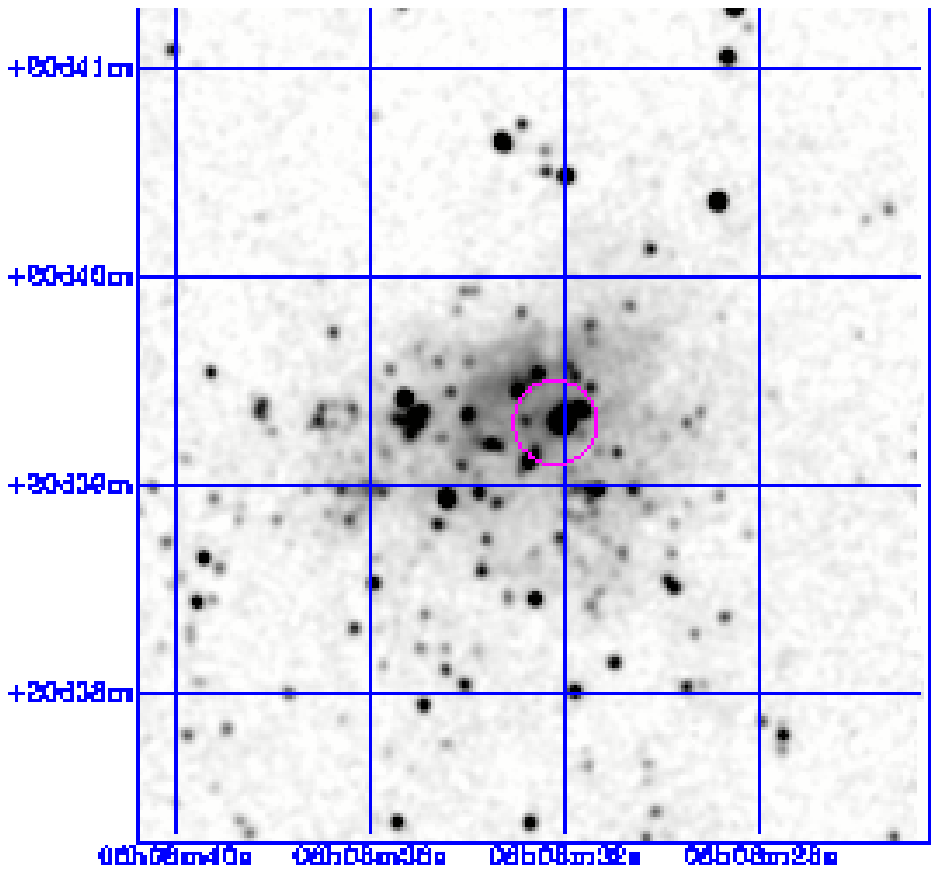}
\end{minipage}\hfill
\begin{minipage}[b]{0.50\linewidth}
\includegraphics[width=\textwidth]{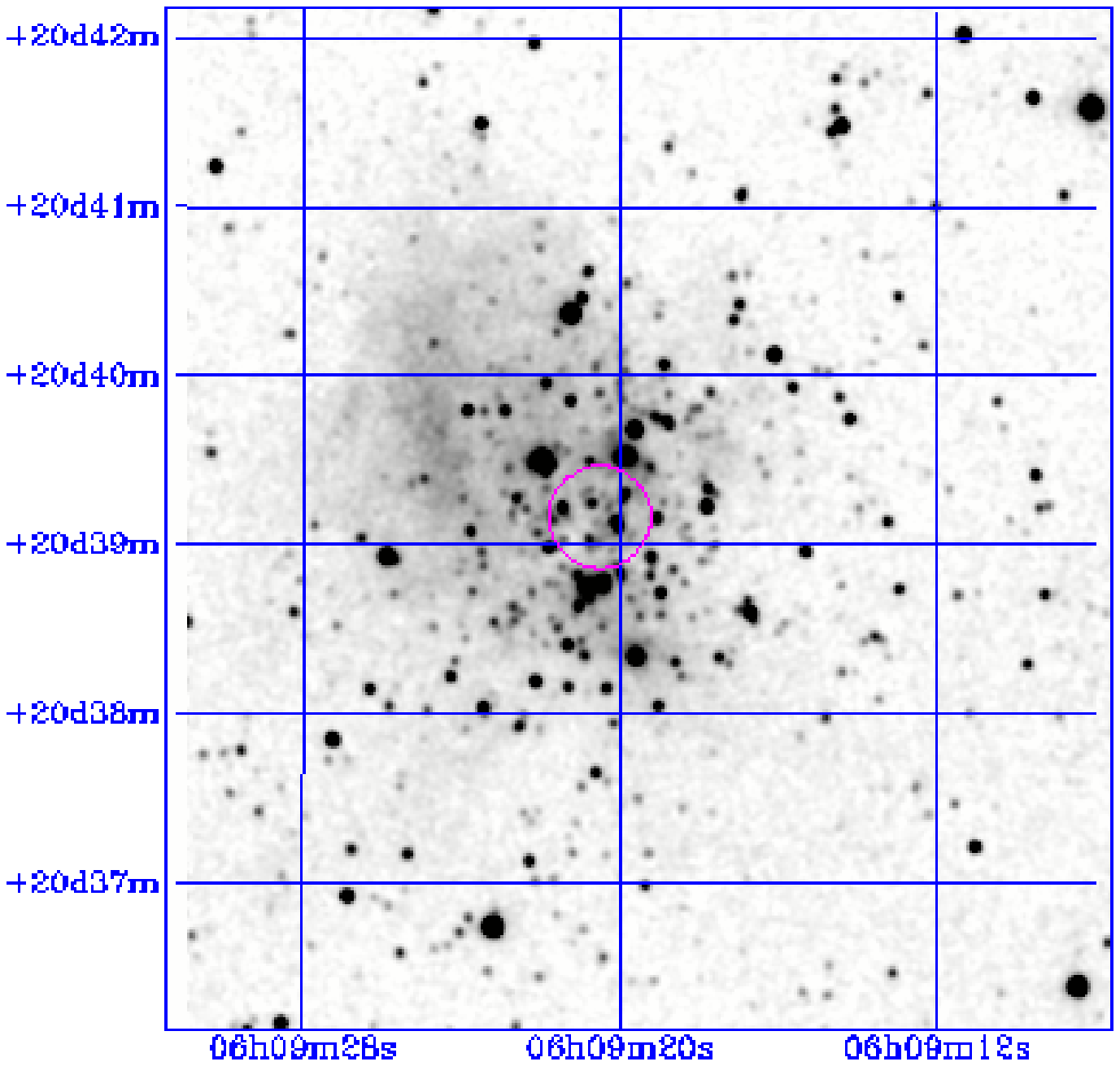}
\end{minipage}\hfill
\begin{minipage}[b]{0.50\linewidth}
\includegraphics[width=\textwidth]{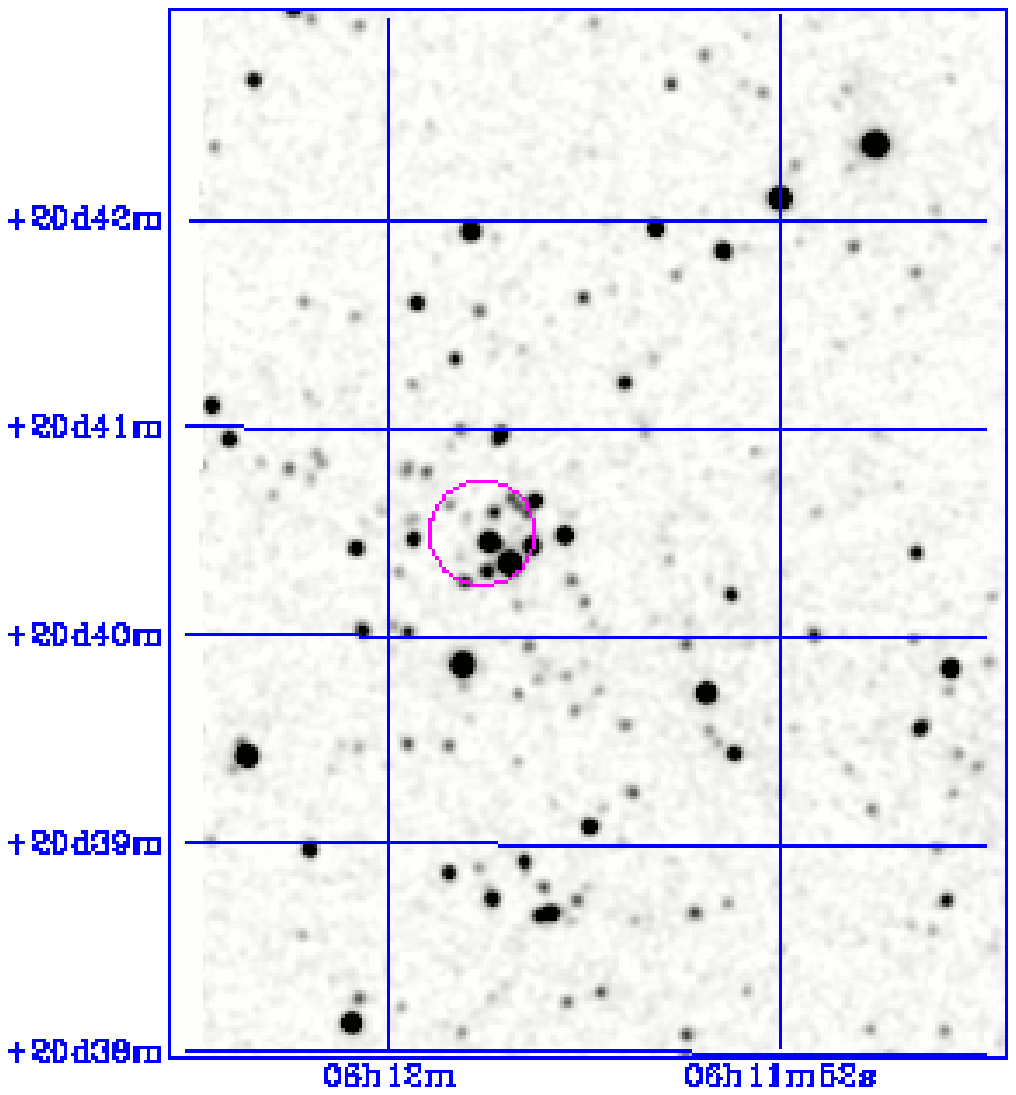}
\end{minipage}\hfill
\begin{minipage}[b]{0.50\linewidth}
\includegraphics[width=\textwidth]{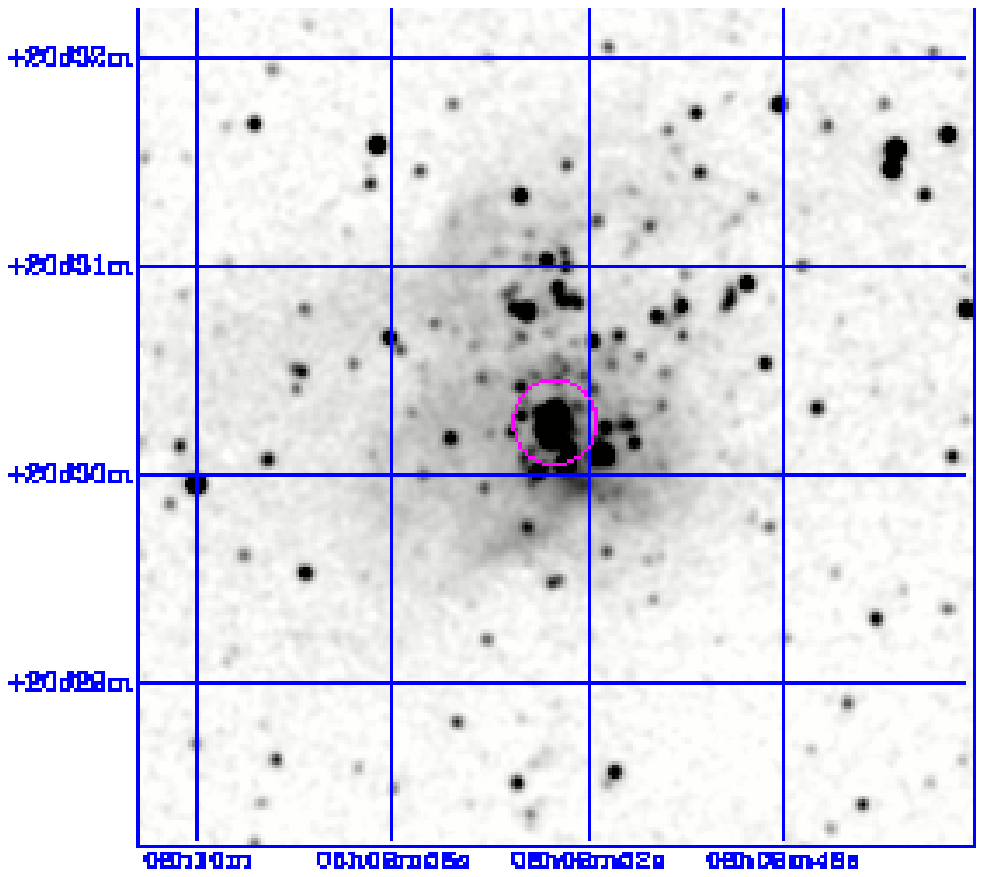}
\end{minipage}\hfill
\begin{minipage}[b]{0.50\linewidth}
\includegraphics[width=\textwidth]{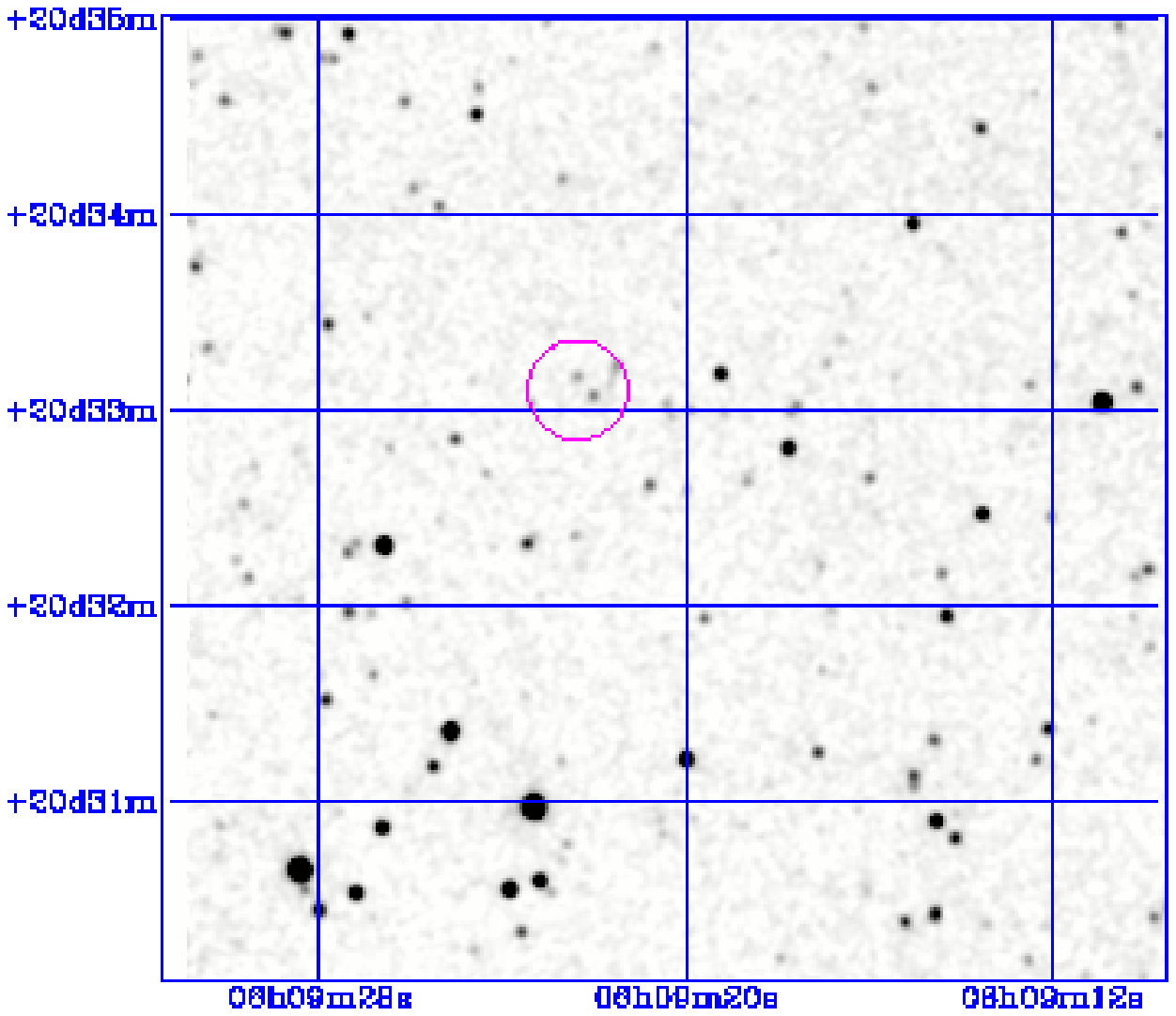}
\end{minipage}\hfill
\caption[]{2MASS \ks\ images of the embedded clusters and the central part of the extended 
stellar group NGC\,2175. From top to bottom and left to right: 
NGC\,2175s ($6\arcmin\times6\arcmin$) and Sh2-252A ($4\arcmin\times4\arcmin$); 
Sh2-252C ($6\arcmin\times6\arcmin$) and Teu\,136 ($5\arcmin\times5\arcmin$); 
Sh2-252E ($4\arcmin\times4\arcmin$) and NGC\,2175 ($6\arcmin\times6\arcmin$).
North to the top and East to the left.}
\label{fig2}
\end{figure}

\begin{table*}
\caption[]{Star clusters or candidates in NGC\,2174 (Sh2-252) identified in the literature}
\label{tab1}
\tiny
\renewcommand{\tabcolsep}{2.0mm}
\renewcommand{\arraystretch}{1.25}
\begin{tabular}{ccccccccc}
\hline\hline
Name&$\ell$&$b$&$\alpha(2000)$&$\delta(2000)$&Size&Other designations&Related objects&References\\
&(\degr)&(\degr)&(hms)&($\degr\,\arcmin\,\arcsec$)&($\arcmin\times\arcmin$)\\
(1)&(2)&(3)&(4)&(5)&(6)&(7)&(8)&(9)\\
\hline
\multicolumn{9}{c}{Embedded clusters}\\
\hline
Sh2-252A & 189.76 & 0.33 & 06:08:32 & +20:39:24 & $1.6\times1.6$ & &
               in NGC\,2175, Sh2-252, Gem\,OB1 & 03,02\\

Sh2-252C & 189.85 & 0.50 & 06:09:22 & +20:39:33 & $2.6\times2.6$ & KKC\,16 &
              in NGC\,2175, Gem\,OB1,rel IRAS\,06063+2040 & 03,02,04\\
              
Sh2-252E & 190.05 & 0.53 & 06:09:53 & +20:30:16 & $1.4\times1.4$ & NGC\,2174,
      CSSS\,16, KKC\,17 & in Gem\,OB1,rel IRAS\,06068+2030 & 01,03,04\\
      
NGC\,2175s$^\dagger$ & 190.07 & 0.79 & 06:10:52 & +20:36:45 & $3\times2$ & OCl-475.1, Pismis\,27 &
       in Sh2-252 & 05,06\\
       
Teutsch\,136 & 190.14 & 1.05 & 06:11:58 & +20:40:28 & $3\times3$ & Koposov\,82& 
       &07,08\\
\hline
\multicolumn{9}{c}{Possible young open cluster or stellar group}\\
\hline
NGC\,2175 & 189.94 & 0.46 & 06:09:22 & +20:33:44 & $22\times15$ & Collinder\,84, OCl-476 &
    in Sh2-252, in Gem\,OB1 & 05\\
\hline
\end{tabular}
\begin{list}{Table Notes.}
\item ($^\dagger$): `s' refers to small; Col.~1: Adopted name; Col.~6: Optical diameter; 
References in Col.~9: 01: \citet{Carpenter93}; 02: these cluster were included in the 
Embedded Cluster Catalogue by \citet{BDB03}; 03: a cluster image is provided in the 
2MASS gallery at {\em http://www.ipac.caltech.edu/2mass/gallery}; 04: \citet{KKC06}; 
05: \citet{Pismis70}; 06: listed in DAML02; 07: \citet{Kron06}; 08: \citet{Koposov08}.
\end{list}
\end{table*}

This paper is organised as follows. In Sect.~\ref{RecAdd} we recall literature data on the targets. 
In Sect.~\ref{2mass} we build the field-star decontaminated CMDs. In Sect.~\ref{struc} we derive 
structural parameters. In Sect.~\ref{N2175} we estimate fundamental parameters for NGC\,2174, 
NGC\,2175, and the individual ECs. In Sect.~\ref{MF} we estimate cluster mass. Concluding remarks 
are given in Sect.~\ref{Conclu}.

\section{Previous data on NGC\,2174 and related objects}
\label{RecAdd}

The prominent H\,II region Sh2-252 was originally catalogued by \cite{Sharp59}, who related 
it to an emission nebula (NGC\,2174) and a loose stellar distribution (NGC\,2175). Currently,
NGC\,2174 refers to the whole H\,II region, while NGC\,2175 refers to a possible young open 
cluster (OC) or a stellar group of scattered stars in the nebula. We will thus refer to 
NGC\,2174 (Sh2-252) for the H\,II region, and NGC\,2175 for the possible stellar group. 

Different designations and parameters for star clusters and candidates in the direction of 
NGC\,2174 are available in the literature. For instance, \citet{Pismis70} and Dias et 
al. (2002, hereafter DAML02 - {\em www.astro.iag.usp.br/$\sim$wilton}) list an OC, NGC\,2175, 
within the H\,II region NGC\,2174. \citet{Pismis70}  finds a small cluster in the area, 
designated as NGC\,2175s, where `s' refers to small. DAML02 refers to NGC\,2175s as Pismis\,27. 
\citet{Pismis70} inferred that NGC\,2175s and NGC\,2175 have different reddening values 
($\ebv=0.70$ and 0.25, respectively), and are located at different distances ($\ds=3.5$ and 
1.95\,kpc, respectively). Star counts also suggest that the distribution of stars in NGC\,2175 
corresponds to a spherical shell. \citet{Pismis70} provides the fundamental 
parameters\footnote{Adopted by WEBDA - {\em www.univie.ac.at/webda}} for the possible OC 
NGC\,2175: $\ds=1.63$\,kpc, $\ebv=0.60$, and age$=9$\,Myr. In DAML02 they are $\ds=1.0$\,kpc, 
$\ebv=0.60$, and age$=32$\,Myr. 
 
For NGC\,2175s, \citet{Koposov08} derived $\ds=1.0$\,kpc, $\ebv=0.68$, and
$\rm age<50$\,Myr, while Teutsch\,136 is cited as a new infrared cluster, but
without parameter determination. \citet{KKC06} derived $\ds=4.52$\,kpc, $A_K=1.0$,
number and mass of member stars $N=202$ and $M=1474\,\ms$, and effective radius
$R_{eff}=2\arcmin$ for Sh2-252C; for Sh2-252E they found $\ds=1.5$\,kpc, $A_K=0.6$,
$N=68$, $M=45\,\ms$, and $R_{eff}=1.5\arcmin$. The remaining objects in the area
(Table~\ref{tab1}) have no parameters yet determined.

Discrepant values for the distance to the H\,II region have also been provided by kinematical 
methods. For instance, based on UBV photometry and spectroscopy, \citet{GGR73} derived a 
kinematical distance of $\ds=1.48\pm1.21$\,kpc. On the other hand, the CO radial velocities 
of \citet{Blitz82} implied the distance $\ds=4.4\pm0.4$\,kpc. More recently, \citet{Reid09}
used trigonometric parallaxes to derive $\ds=2.1$\,kpc and a (revised) kinematical distance
of $\ds=3.3^{+4.2}_{-2.4}$\,kpc. Such a difference in the kinematical distance may be accounted 
for by the nearly anti-Galactic direction of Sh2-252.

\begin{table*}
\caption[]{Fundamental parameters derived in this work}
\label{tab2}
\renewcommand{\tabcolsep}{3.9mm}
\renewcommand{\arraystretch}{1.25}
\begin{tabular}{ccccccccc}
\hline\hline
Cluster&$\alpha(2000)$&$\delta(2000)$&$\ell$&$b$&Age&\ebv&\ds&\dSC\\
 & (hms)&($\degr\,\arcmin\,\arcsec$)&(\degr)&(\degr)&(Myr)&(mag)&(kpc)&(kpc)\\
(1)&(2)&(3)&(4)&(5)&(6)&(7)&(8)&(9)\\
\hline
\multicolumn{9}{c}{Embedded clusters}\\
\hline
Sh2-252A&06:08:32.2&$+$20:39:18.0&189.77&$+$0.34&$\sim5$&$0.29\pm0.16$&$1.4\pm0.3$&$1.4\pm0.5$\\

Sh2-252C&06:09:21.6&$+$20:38:37.0&189.87&$+$0.50&$\sim5$&$0.30\pm0.16$&$1.5\pm0.4$&$1.5\pm0.5$\\

Sh2-252E&06:09:52.7&$+$20:30:15.2&190.05&$+$0.54&$\sim5$&$0.32\pm0.16$&$1.4\pm0.3$&$1.4\pm0.4$\\

NGC\,2175s&06:10:54.6&$+$20:36:49.5&190.07&$+$0.80&$\sim5$&$0.45\pm0.10$&$1.2\pm0.3$&$1.1\pm0.3$\\

Teutsch\,136&06:11:58.1&$+$20:49:29.5&190.14&$+$1.05&$\sim5$&$0.54\pm0.16$&$1.8\pm0.4$&$1.7\pm0.5$\\
\hline
\multicolumn{9}{c}{Possible young open cluster or stellar group}\\
\hline
NGC\,2175&06:09:22.4&$+$20:33:06.5&189.95&$+$0.46&$\sim5$&$0.35\pm0.10$&$1.5\pm0.4$&$1.7\pm0.5$\\
         
\hline
\end{tabular}
\begin{list}{Table Notes.}
\item Cols.~4 and 5: Galactic coordinates; Col.~8: distance from the Sun; col.~9: distance 
from the Solar circle.
\end{list}
\end{table*}

A few relatively bright stars mixed with nebular gas and/or dust emission are seen in the
$45\arcmin\times45\arcmin$ B image (Fig.~\ref{fig1}, taken from LEDAS\footnote{Leicester 
Database and Archive Service (LEDAS) DSS/DSS-II service on ALBION; 
{\em ledas-www.star.le.ac.uk/DSSimage}.}). Close-ups of the clusters (and the stellar group 
NGC\,2175) are shown in the smaller field 2MASS \ks\ images. Table~\ref{tab1} provides parameters 
found in the literature for the clusters and candidates. Cluster designations and coordinates - as 
derived in the present paper - are given in Table~\ref{tab2}. Note the small differences between 
our coordinates and those given in the literature. This occurs because the central coordinates were 
re-computed with field-star decontaminated photometry (Sect.~\ref{2mass}). According to our approach, 
the cluster centre corresponds to the coordinates that produce the smoothest stellar radial density 
profile (RDP) and, at the same time, the highest stellar density in the innermost region (Sect.~\ref{struc}). 

\section{Decontaminated CMDs}
\label{2mass}

Figure~\ref{fig1} shows that the H\,II region NGC\,2174 (and related objects) still retain 
part of the parent gas and dust. {\bf Thus, we employ the 2MASS \jj, \hh, and \ks\ 
photometry to probe the photometric properties with an adequate depth, especially at the 
faint stellar sequences. Additionally, 2MASS provides the spatial and photometric uniformity 
required for wide extractions and high star-count statistics. The photometry for each target 
(Table~\ref{tab1}) was extracted from 
VizieR\footnote{\em http://vizier.u-strasbg.fr/viz-bin/VizieR?-source=II/246},
and only stars with errors lower than 0.1\,mag are used. 

Clusters that contain important fractions of faint stars and/or are projected near the 
Galactic equator, require field-star decontamination for the proper identification 
and characterisation of the member stars. Although projected towards the Galactic anti-centre, 
the present clusters are found near the plane, having CMDs dominated by PMS stars
(see below). Thus, we apply the decontamination algorithm first developed in \citet{BB07} 
and improved in \citet{vdB92} to minimise confusion with red dwarfs of the Galactic field.

Representative CMDs of the targets are shown in the top panels of Figs.~\ref{fig3} and 
\ref{fig4}. For NGC\,2175, we first consider a region of radius $R=5\arcmin$, which is 
located outside the borders of the neighbouring Sh2-252C and Sh2-252E (Fig.~\ref{fig1})
and, thus, is expected to be free from contamination by stars of both objects; a wider 
region will be considered in Sect.~\ref{N2175}. An indication of a young age
comes from a comparison with the CMDs extracted from equal-area offset fields (middle 
panels), which is consistent with the presence of gas and dust (Figs.~\ref{fig1} and 
\ref{fig2}).

\begin{figure}
\resizebox{\hsize}{!}{\includegraphics{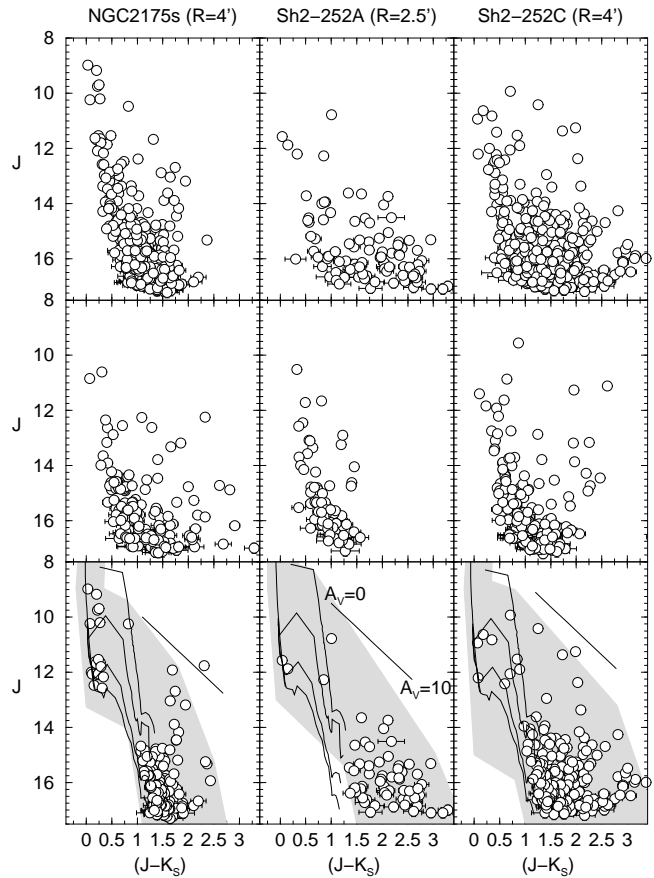}}
\caption[]{CMDs showing the observed photometry for representative regions 
(top) and the equal-area comparison fields (middle). The decontaminated CMDs are shown in 
the bottom panels, together with the 5\,Myr Padova isochrone (for the MS) and the 0.2, 1,
5, and 10\,Myr PMS isochrones. The isochrones have been set according to the adopted
fundamental parameters (Sect.~\ref{DFP2}). Also shown is the colour-magnitude filter 
(shaded polygon). Reddening vectors for $\aV=0-10$ are shown in the bottom panels.}
\label{fig3}
\end{figure}

\begin{figure}
\resizebox{\hsize}{!}{\includegraphics{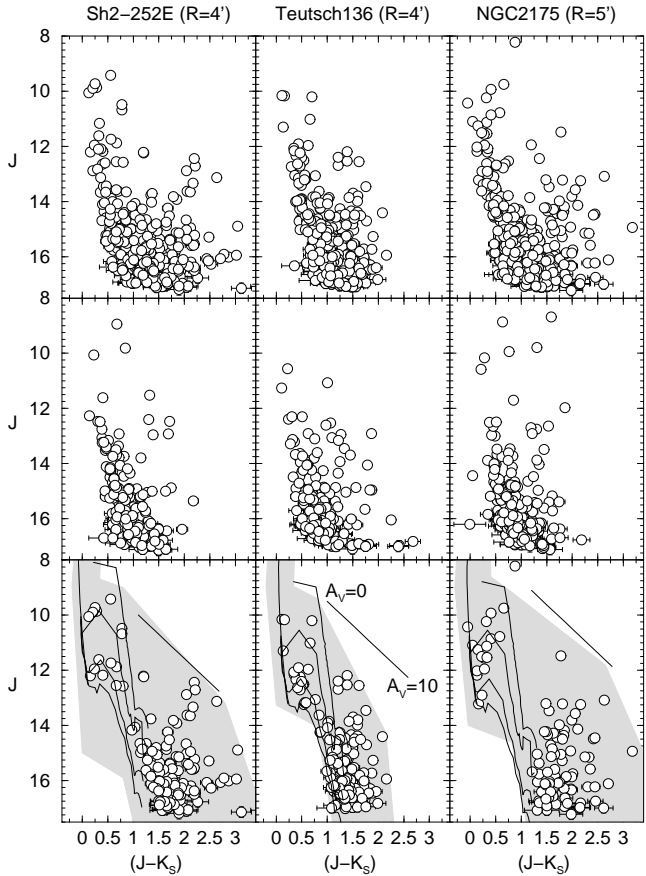}}
\caption[]{Same as Fig.~\ref{fig3} for the remaining CMDs. The extraction for the stellar
group NGC\,2175 corresponds to a region free from contamination by neighbouring objects.}
\label{fig4}
\end{figure}

Indeed, the decontaminated CMDs (bottom panels of Figs.~\ref{fig3} and \ref{fig4}) present 
similar features: they are essentially characterised by stellar sequences of mildly reddened,
young clusters, with a nearly-vertical, developing MS and a significant population of PMS stars. }
Given the time-scales associated with stellar formation ($\sim10^7$\,yr for low-mass stars), 
very young clusters are expected to contain a population of PMS stars (e.g. \citealt{vdB92}, 
and references therein). Thus, the assumption that the red and faint stars belong to the PMS 
is consistent with the $\sim5$\,Myr of age of the ECs in the complex (Sect.~\ref{N2175}). 
Internal differential reddening is implied by the colour distribution at faint magnitudes 
($\jj\ga14$), which is wider than the spread predicted purely by PMS models. A comparison with 
the reddening vector (for $\aV=0~{\rm to}~10$) shows different degrees of differential reddening, 
being lower for NGC\,2175s and Teu\,136, and higher for the remaining cases. If most of the colour 
spread is due to non-uniform reddening - and not to systematic differences in the stellar 
content - the upper limit to the differential reddening would be $\Delta\aV\la6$\,mag.

\section{Structural parameters}
\label{struc}

{\bf Probable member stars of each cluster are selected by a colour-magnitude filter, which 
is wide enough to include MS and PMS stars, the photometric uncertainties, and binaries 
(Figs.~\ref{fig3} and \ref{fig4}); they are used to build the projected stellar radial density 
profile (RDP) around the cluster centre.} As discussed elsewhere (e.g. \citealt{vdB92}, and 
references therein), this filtering enhances the RDP contrast relative to the background, 
especially in crowded fields (e.g. \citealt{BB07}). 

\begin{figure}
\resizebox{\hsize}{!}{\includegraphics{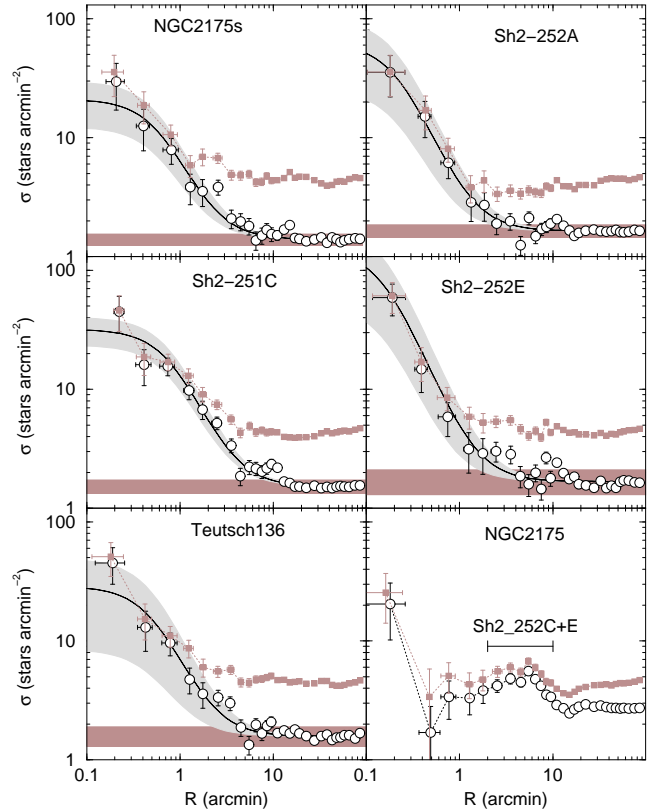}}
\caption[]{Colour-magnitude filtered RDPs (circles) built for the MS and PMS stars combined, 
together with the best-fit King-like profile (solid line), the $1\sigma$ uncertainty 
(light-shaded region) and the background level (heavy-shaded polygon). For comparison, the
RDPs built with the observed photometry (squares) are also shown.}
\label{fig5}
\end{figure}

Both sets of RDPs (before and after applying the colour-magnitude filters) are shown in 
Fig.~\ref{fig5}, where the contrast enhancement due to the filtering is clearly seen. 
Except for NGC\,2175, the remaining RDPs are typical of star clusters projected on a
stellar fore/background: they follow a power-law (on a log-log scale) with a core near 
the centre, usually with fluctuations due to neighbouring clusters and/or stellar 
concentrations. {\bf Indeed, the RDPs are well represented by the King-like (\citealt{King1962}) 
function $\sigma(R)=\sigma_{bg}+\sigma_0/(1+(R/R_c)^2)$, where $\sigma_0$ and $\sigma_{bg}$ 
are the central and residual stellar densities, and \rc\ is the core radius.} 


We also use the RDPs to estimate the cluster radius (\rl), taken as the distance from the 
cluster centre where the RDP and field fluctuations are statistically indistinguishable 
(e.g. \citealt{DetAnalOCs}). In this sense, \rl\ is an observational truncation 
radius, whose value depends both on the radial distribution of member stars and the field 
density. The RDP fits (and uncertainties) are shown in Fig.~\ref{fig5}, and the 
corresponding structural parameters are given in Table~\ref{tab3}. 

\begin{table*}
\caption[]{Structural parameters derived for the embedded clusters}
\label{tab3}
\renewcommand{\tabcolsep}{3.0mm}
\renewcommand{\arraystretch}{1.25}
\begin{tabular}{ccccccccccc}
\hline\hline
Cluster&$\sigma_0$&\rc&\rl&&$1\,arcmin$&&$\sigma_0$&\rc&\rl\\
       &$\rm(stars~arcmin^{-2})$&(arcmin)&(arcmin)&&(pc)&&$\rm(stars~pc^{-2})$&(pc)&(pc)\\
(1)&(2)&(3)&(4)&&(5)&&(6)&(7)&(8)\\
\hline
Pismis\,27&$19.6\pm8.7$&$0.58\pm0.18$&$5.5\pm0.5$&&0.335&&$175\pm77$&$0.19\pm0.06$&$1.8\pm0.2$\\

Sh2-252\,A&$60.8\pm37.3$&$0.21\pm0.08$&$2.4\pm0.4$&&0.410&&$362\pm222$&$0.09\pm0.03$&$1.0\pm0.2$\\

Sh2-252\,C&$30.3\pm8.4$&$0.78\pm0.14$&$4.5\pm0.5$&&0.447&&$152\pm42$&$0.35\pm0.06$&$2.0\pm0.2$\\

NGC\,2174&$190\pm100$&$0.15\pm0.10$&$4.0\pm0.5$&&0.405&&$916\pm610$&$0.06\pm0.04$&$1.6\pm0.2$\\

Teutsch\,136&$27.1\pm20.2$&$0.50\pm0.25$&$4.5\pm0.5$&&0.511&&$104\pm77$&$0.26\pm0.13$&$2.3\pm0.3$\\
   
\hline
\end{tabular}
\begin{list}{Table Notes.}
\item Col.~5: arcmin to parsec scale.
\end{list}
\end{table*} 


{\bf The RDPs are essentially defined by the red and faint stars (that we identify as PMS), 
since they predominate in the CMDs by $\ga95\%$ in number. Thus, coupled to the field-star 
decontamination, the cluster-like RDPs consistently indicate that these stars belong to the 
PMS, not being unsubtracted red dwarfs. Otherwise, the RDPs would be characterised by random fluctuations (e.g. \citealt{AntiC}).} 


The values of \rc\ and \rl\ (Table~\ref{tab3}) show that we are dealing with
small-scale clusters. Indeed, compared to the core radii derived for a sample of relatively 
nearby OCs by \citet{Piskunov07}, the present ECs occupy the small-\rc\ tail of the
distribution.

\section{The distance to NGC\,2174 and the nature of NGC\,2175}
\label{N2175}

\subsection{Distance and age of NGC\,2174}
\label{dist2174}

Evidence drawn in previous sections suggest that we are dealing with a coeval star formation in 
NGC\,2174 that produced at least 5 ECs physically related to the H\,II region. In this sense, we 
can use the CMD morphology of the whole area to derive the average fundamental parameters of 
NGC\,2174. For this purpose, {\bf we decontaminate a region of 25\arcmin\ in radius 
(Fig.~\ref{fig6}), thus including NGC\,2175 and the ECs Sh2-252A, C, E, and NGC\,2175s. }

To derive the fundamental parameters we use the Padova isochrones (\citealt{Girardi2002}) computed 
for the 2MASS filters\footnote{{\em http://stev.oapd.inaf.it/cgi-bin/cmd}}. For the PMS we use the 
isochrones of \citet{Siess2000}. We restrict the analysis to solar metallicity isochrones because 
the clusters are expected to be young and located not far from the Solar circle (see below), a region
essentially occupied by $[Fe/H]\approx0.0$ OCs (\citealt{Friel95}). Reddening transformations are 
based on the absorption relations $A_J/A_V=0.276$, $A_H/A_V=0.176$, $A_{K_S}/A_V=0.118$, and
$A_J=2.76\times\ejh$ (\citealt{DSB2002}), with $R_V=3.1$, considering the extinction curve of
\citet{Cardelli89}. 


{\bf Given the poorly-populated MS, the significant population of PMS stars, and the differential 
reddening, we estimate the fundamental parameters {\em by eye}, using the decontaminated CMD
morphology (Fig.~\ref{fig6}) as constraint. Beginning with zero distance modulus and reddening, 
the MS$+$PMS isochrones are shifted in colour and magnitude until an acceptable fit for the blue 
border of the MS and PMS sequences is obtained. The rather poorly-populated and nearly vertical 
MS accepts any isochrone of age within 1---10\,Myr. Considering the differential reddening, a 
similar age spread results for the PMS stars, which are basically contained within the 0.2\,Myr 
and 10\,Myr isochrones. This age range is consistent with the gas and dust-embedded ECs in the 
area, suggesting an age-spread of $\sim10$\,Myr in the star formation.

\begin{figure}
\resizebox{\hsize}{!}{\includegraphics{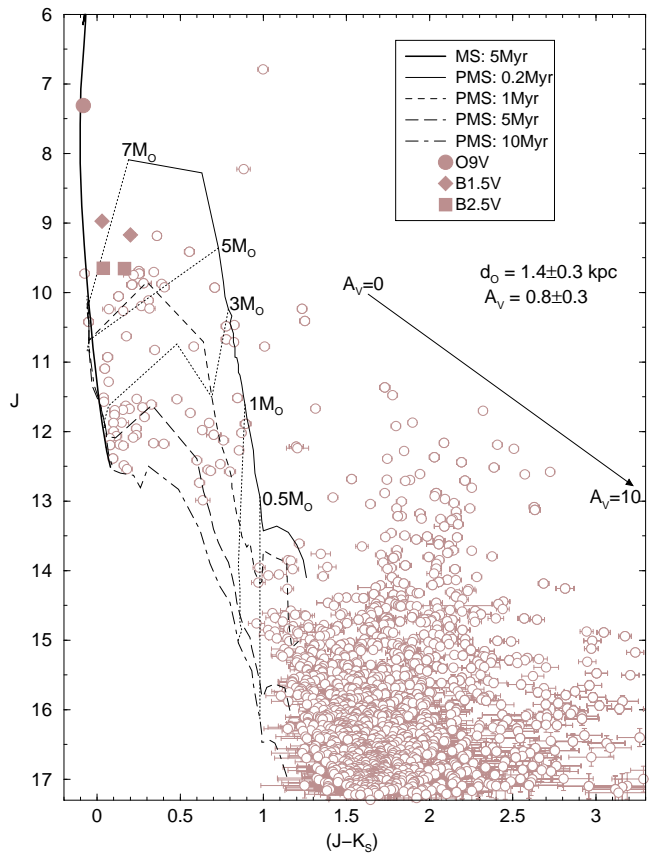}}
\caption[]{Decontaminated photometry of NGC\,2174, covering a region of radius $R=25\arcmin$. 
The isochrones have been set with $\aV=0.8\pm0.3$ and $\ds=1.44\pm0.34$\,kpc. Evolutive tracks
of selected PMS masses are shown (light-dotted lines) for illustrative purposes. Filled symbols
indicate MS O and B stars in the area.}
\label{fig6}
\end{figure}

Additionally, the presence of a few O and B stars still in the MS, some of these projected 
within the ECs Sh2-252E and NGC\,2175s, constrain the age of the complex to a few Myr. The 
spectral types, coordinates (J2000), and location of these ionising stars are: O9\,V 
($\alpha$=06:09:40, $\delta$=+20:29:15.4), near the west border of Sh2-252E; B1.5\,V
($\alpha$=06:10:53, $\delta$=+20:36:33.8), near the centre of NGC\,2175s; B1.5\,V ($\alpha$=06:10:59, 
$\delta$=+20:34:19.6), slightly to the south of the latter; and two B2.5\,V stars 
($\alpha$=06:09:50, $\delta$=+20:37:05.2, and $\alpha$=06:09:55, $\delta$=+20:38:31.6), located to 
the north-east of the centre of NGC\,2175. Thus, based on the gas and dust-embedded character of the 
ECs (Figs.~\ref{fig1} and \ref{fig2}), and the ionising stars, we adopt $\sim5$\,Myr as the age for 
the bulk of stars in NGC\,2174. 

The adopted solution is shown in Fig.~\ref{fig6}, where the isochrones are set with $\aV=0.8\pm0.3$ 
and $\ds=1.44\pm0.34$\,kpc. Within the uncertainty, this value is consistent with the distance of 
the Gem\,OB1 association ($\ds=1.5-1.9$\,kpc), to which NGC\,2174 appears to be physically related 
(\citealt{Dunham2010}). }

\subsection{Nature of NGC\,2175}
\label{Nat2175}

{\bf According to the literature (Table~\ref{tab1}), NGC\,2175 encompasses the 
ECs Sh2-252C and Sh2-252E (Fig.~\ref{fig1}). In fact, the decontamination shows that most of 
the stars ($\approx64\%$) in the region belong to both ECs, with the remaining following 
the same CMD morphology as the ECs (Fig.~\ref{fig4}), which might suggest an EC. However, the 
RDP centred on NGC\,2175 (Fig.~\ref{fig5}) is irregular and shows only a very narrow peak, 
followed by the excesses produced by the neighbouring ECs. This indicates that NGC\,2175 
is not a star cluster. Instead, its stellar content may be {\em (i)} stars that already have 
escaped from the neighbouring ECs, {\em (ii)} a leftover of a disrupted EC, or {\em (iii)} a 
result of a star formation throughout the molecular cloud not related to a cluster. Perhaps, 
given the rather young age of the star-formation event, the first two scenarios may not apply, 
since the stars would have to scatter over a region of $\sim6$\,pc in radius in less than
$\sim5-10$\,Myr. This would favour NGC\,2175 as an off-cluster star formation in the molecular 
cloud. }

\subsection{Fundamental parameters of individual clusters}
\label{DFP2}

Since the decontaminated CMD morphologies of the individual clusters (Figs.~\ref{fig3} and 
\ref{fig4}) resemble that of the wide-field CMD of NGC\,2174 (Fig.~\ref{fig6}), a similar 
isochrone solution is expected to apply to all ECs in the complex. Thus, we simply use the 
NGC\,2174 solution as a fundamental-parameter template. The adopted solutions are shown in 
Figs.~\ref{fig3} and \ref{fig4}, and the fundamental parameters are given in Table~\ref{tab2}. 

We find only small variations on reddening ($\ebv=0.29 - 0.45$) and distance 
($\ds=1.2 - 1.5$\,kpc) among the CMDs (Table~\ref{tab2}). Teu\,136 may be the exception, 
in the sense that it is somewhat more reddened ($\ebv=0.54$) and more distant
($\ds=1.8\pm0.4$\,kpc) than the others. Obviously, given the uncertainty in \ds, 
the clusters can be said to be at the same distance. However, since Teu\,136 is located beyond 
the east border of the H\,II region (Fig.~\ref{fig1}), it may be just a young cluster projected 
close, but not physically related to the complex. Our reddening value for Sh2-252C is about 
10\% of that derived by \citet{KKC06}, while our distance from the Sun is about 30\% of theirs. 
On the other hand, both works agree with respect to the distance of Sh2-252E. 

Compared to the remaining cases (Figs.~\ref{fig3} and \ref{fig4}), the decontaminated CMD of the 
dust-shrouded EC NGC\,2175s presents the best constraints, in terms of stellar sequences (it 
contains at least 2 B1.5\,V stars), for finding an independent isochrone solution. Under the same 
strategy applied for NGC\,2174, we find $\ejh=0.14\pm0.03$ ($\ebv=0.45\pm0.10$ or $A_V=1.4\pm0.3$), 
the observed and absolute distance moduli $\mMJ=10.7\pm0.5$ and $\mMo=10.31\pm0.51$, respectively,
and $\ds=1.2\pm0.3$\,kpc. These values are consistent with those of the wide-field NGC\,2174,
and with the $\ds=1$\,kpc obtained by \citet{Koposov08}. Given the $\ell$ and $b$ coordinates
(Table~\ref{tab1}), NGC\,2175s is located $\approx1.1$\,kpc outside the Solar circle and
$\approx16$\,pc above the plane. 

\section{Stellar mass estimate}
\label{MF}

Followed by a developing, poorly-populated MS, PMS stars are the dominant (in number)
component in the CMDs of our clusters (Figs.~\ref{fig3} and \ref{fig4}). Consequently, 
most of the cluster mass is still stored in the PMS stars. Thus, to estimate the cluster 
mass we simply consider the number of MS and PMS stars (for $R\le\rl$) on the 
field-decontaminated photometry. 

The mass of each MS star is taken from the mass-luminosity relation corresponding to the 
isochrone solution (Sect.~\ref{DFP2}). Summing up these values for all stars produces the 
total number ($n_{MS}$) and mass ($m_{MS}$) of MS stars. MS stars are detected within
the range $1.7 - 6.5\,\ms$ (Table~\ref{tab4}). For the PMS, on the other hand, 
the presence of differential reddening precludes such an estimate of individual masses. 
Thus, we simply count the number of PMS stars and adopt an average PMS mass value to 
estimate $n_{PMS}$ and $m_{PMS}$. To compute the average
PMS mass value we use the \citet{Kroupa2001} initial mass function\footnote{Defined
as $dN/dM\propto m^{-(1+\chi)}$, it assumes the slopes $\chi=0.3\pm0.5$ for 
$0.08\leq m(\ms)\leq0.5$, $\chi=1.3\pm0.3$ for $0.5\leq m(\ms)\leq1.0$, and 
$\chi=1.35$ for $m(\ms)>1.0$.} for PMS masses between $0.08\,\ms - 7\,\ms$. The
result is $\bar{m}_{PMS}=0.6\,\ms$. The estimated values are given in Table~\ref{tab4},
which also gives the average mass density ($\rho$) of each cluster.

\begin{table*}
\caption[]{MS and PMS stellar content derived for the embedded clusters}
\label{tab4}
\renewcommand{\tabcolsep}{3.2mm}
\renewcommand{\arraystretch}{1.25}
\begin{tabular}{ccccccccccc}
\hline\hline
&\multicolumn{3}{c}{MS}&&\multicolumn{2}{c}{PMS}&&\multicolumn{3}{c}{MS$+$PMS}\\
\cline{2-4}\cline{6-7}\cline{9-11}
Cluster&$\Delta\,m$&$n_{MS}$&$m_{MS}$&&$n_{PMS}$&$m_{PMS}$&&$n_{MS+PMS}$&$m_{MS+PMS}$&$\rho$\\
&(\ms)&(stars)&(\ms)&&(stars)&(\ms)&&(stars)&(\ms)&($\ms\,pc^{-3}$)\\
\hline
(1)&(2)&(3)&(4)&&(5)&(6)&&(7)&(8)&(9)\\
\hline
NGC\,2175s&1.7-6.5&$18\pm4$&$51\pm12$&&$188\pm14$&$113\pm8$&&$196\pm18$&$164\pm14$&$6.7\pm0.6$\\
   
Sh2-252A &2.7-3.8&$2\pm1$&$6\pm3$   &&$86\pm8$  &$52\pm5$&&$88\pm10$&$58\pm6$&$13.8\pm1.4$\\
   
Sh2-252C &1.7-6.5&$6\pm2$&$20\pm9$  &&$289\pm15$&$173\pm9$&&$295\pm17$&$193\pm13$&$5.8\pm0.4$\\

Sh2-252E &1.7-5.5&$6\pm2$&$20\pm9$  &&$109\pm12$&$65\pm7$&&$115\pm13$&$85\pm11$&$4.9\pm0.7$\\

Teu\,136  &2.7-5.5&$4\pm2$&$16\pm8$  &&$176\pm14$&$106\pm8$&&$180\pm15$&$122\pm11$&$2.4\pm0.2$\\

\hline
\end{tabular}
\begin{list}{Table Notes.}
\item Col.~2: Effective mass range of the MS. Col.~9: average mass density.
\end{list}
\end{table*}

As anticipated by the CMDs (Figs.~\ref{fig3} and \ref{fig4}) and the small sizes 
(Table~\ref{tab3}), we are dealing with low-mass clusters ($\la200\,\ms$) having 
poorly-populated MSs and with most ($\ga95\%$) of the stars still in the PMS. Indeed, 
the mass stored in the PMS stars is the dominant ($\approx70\%$ to $\approx90\%$) 
component of the detected cluster mass, which is consistent with the young ages. However, 
given the presence of dust and gas (Figs.~\ref{fig1} and \ref{fig2}), differential
reddening, and the 2MASS photometric limitation (which precludes detection of 
very-low mass PMS stars), the mass values may be somewhat higher than quoted in 
Table~\ref{tab4}.

\section{Summary and conclusions}
\label{Conclu}

Previous works on the H\,II region NGC\,2174 (Sh2-252) came up with discrepant values 
for the age and distance from the Sun of the deeply-embedded star clusters (and 
candidates) in the area, to the point that no physical relation among them - and the 
complex - could have been inferred.

We investigate the above issue with field-star decontaminated 2MASS photometry (to enhance 
CMD evolutionary sequences) and stellar RDPs, to derive fundamental and structural parameters 
of the 5 previously catalogued embedded clusters (Sh2-252A, C, E, NGC\,2175s, and Teu\,136) 
and one candidate (NGC\,2175).

The decontaminated CMDs are characterised by similar properties: a poorly-populated and 
developing MS, a dominant fraction ($\ga95\%$ in number) of PMS stars, a similar foreground 
absorption, $\aV\sim1$\,mag, and some differential reddening. Taken together, the presence 
of gas, dust, some O\,V and B\,V stars, and the MS$+$PMS CMD morphologies consistently 
constrain the age of the ECs (and the extended stellar group) to less than 
$\approx10$\,Myr (the bulk of the stars are probably $\sim5$\,Myr old), with a time-spread 
of $\sim10$\,Myr for the star formation. The MS$+$PMS stellar masses are low, within 
$\approx60\,\ms$ to $\approx200\,\ms$. Within the uncertainties, the distances from 
the Sun of Sh2-252A, C, E, NGC\,2175s, and NGC\,2175 are essentially the same, 
$\ds\approx1.4\pm0.4$\,kpc, which agrees with that estimated for the physically-related 
association Gem\,OB1. Teu\,136 appears to be slightly more distant ($\ds=1.8\pm0.4$\,kpc) 
and reddened ($\aV\sim1.8$\,mag) than the other objects. Since Teu\,136 is beyond the 
north-east border of NGC\,2174, it may be just an EC projected near the H\,II region. The 
decontaminated wide-field CMD of NGC\,2174, which is expected to reflect the average 
properties of the stars in the region, provides the distance $\ds=1.44\pm0.34$\,kpc and 
the foreground absorption $\aV=0.8\pm0.3$.

The stellar RDPs of Sh2-252A, C, E, NGC\,2175s, and Teu\,136 follow a King-like function 
characterised by small core and cluster radii, with $\rm0.06\la\rc(pc)\la0.26$ and 
$\rm1.0\la\rl(pc)\la2.3$, respectively. NGC\,2175, on the other hand, is not a cluster, and 
its stars probably originated on the same star-formation event that gave rise to the ECs. Thus, 
NGC\,2175 may be classified as a young stellar group.

What can be concluded from our uniform, near-infrared approach, is that NGC\,2174 and its 
related ECs (Sh2-252A, C, E, NGC\,2175s, and perhaps, Teu\,136) are  part of a single 
star-forming complex located at $\ds\approx1.4$\,kpc from the Sun. Thus, we are dealing with 
a coeval (to within $\pm5$\,Myr) star-forming event that extended for $\sim10$\,Myr. Besides 
the scattered stars of NGC\,2175 (probably an off-cluster star formation in the molecular 
cloud), the event gave rise to at least 4 ECs in the complex. Finally, the derivation of 
constrained distance values for star-forming complexes is important for spiral arm structure 
studies (e.g. \citealt{Russeil03}), by providing a kinematically-independent determination, 
especially for central and anti-centre directions.


\section*{Acknowledgements}
{\bf We thank an anonymous referee for interesting comments and suggestions.}
We acknowledge support from the Brazilian Institution CNPq.
This publication makes use of data products from the Two Micron All Sky Survey, which
is a joint project of the University of Massachusetts and the Infrared Processing and
Analysis Centre/California Institute of Technology, funded by the National Aeronautics
and Space Administration and the National Science Foundation. This research has made 
use of the WEBDA database, operated at the Institute for Astronomy of the University
of Vienna.

\label{lastpage}
\end{document}